\documentclass[twocolumn,prl,superscriptaddress,floatfix]{revtex4}
\usepackage{graphicx}
\usepackage{xcolor}
\usepackage{epsfig}
\usepackage{rotating}
\usepackage{amsmath}
\usepackage{amsfonts}
\usepackage{amssymb}
\usepackage{enumerate}
\usepackage{longtable}
\usepackage{appendix}
\usepackage{dcolumn}
\usepackage{bm}

\usepackage{hyperref}
\usepackage{ulem}

\begin{document}

\title{Onset of charge incompressibility and Mott gaps in the Honeycomb-Lattice SU(4) Hubbard Model: Lessons for Twisted Bilayer Graphene systems}

\author{Rahul Hingorani}
\affiliation{Department of Physics, University of California Davis, 
CA 95616, USA}

\author{Jaan Oitmaa}
\affiliation{School of Physics, The University of New South Wales,
Sydney 2052, Australia}

\author{Rajiv R. P. Singh}
\affiliation{Department of Physics, University of California Davis, 
CA 95616, USA}

\date{\rm\today}

\begin{abstract}

We use finite temperature strong coupling expansions to calculate thermodynamic properties
of the Honeycomb-lattice SU(4) Hubbard model. We present numerical results for various properties including
chemical potential, compressibility, entropy and specific heat as a function of temperature and density at several $U/t$ values. 
We study the onset of charge incompressibility and Mott gaps  as the temperature is lowered at integer densities. 
In the incompressible Mott regime, the expansions are recast into a high temperature expansion for a generalized spin model with SU(4) symmetry, which is then used to study the convergence of strong coupling expansions in t/U.
We discuss lessons that can be drawn from high temperature properties of a simple Hubbard model regarding Twisted Bilayer Graphene (TBG) and other magic-angle flat-band systems. 

\end{abstract}


\maketitle

{\bf Introduction:}
Recent years have seen increasing interest in the Fermionic Hubbard model with more than two spin species
\cite{bloch,honerkamp,taie,lorenzo2,lorenzo,padilla,yoshida,wu}. In cold atom platforms, atoms can have internal degrees of freedom such as those associated with hyperfine states. Contact interactions between atoms  provide excellent realizations of local Hubbard interaction $U$ with SU(N) symmetry. Even in the solid state context, SU(4) symmetry can arise for multi-orbital systems although full SU(4) symmetry generically requires fine-tuning of parameters \cite{khomskii,yamada}. A recent realization of systems with potential SU(4) symmetry are Twisted Multilayer systems at magic angles with extremely narrow or nearly flat bands \cite{MB, ashvin-SU4,fu1,fu2,fu3,vafek,TLG-Hubbard,ashvin-SUN,zhang-mao}. Flat bands can provide both strong correlations and enhanced internal symmetry. While the low temperature phase behavior of these systems clearly depend on the complex band structure
and details which break SU(N) symmetry \cite{zhang-mao}, 
a question of interest in this work is: What lessons can be drawn by comparison with high temperature thermodynamics  of a simple SU(N) Hubbard model?

Strong coupling expansions around the atomic limit provide a powerful formalism to calculate temperature dependent properties of Hubbard models in the thermodynamic limit \cite{oitmaa-book, oitmaa2, ten-haaf,rrps1,rrps2}. These expansions can be developed at inverse temperature temperature $\beta$ and chemical potential $\mu$ in the grand canonical ensemble and for any set of hopping parameters.
Each term in the expansion depends on 
$\beta t$, the fugacity $\zeta=\exp{\beta \mu}$, and the Hubbard $U$ which enters the expansions both in terms of $w=\exp{-\beta U}$ and $1/\beta U$. 

One of our main focus in this paper is the temperature dependence of the electronic  compressibility and entropy. As the temperature is lowered below $U$, at most densities the compressibility becomes large while at integer densities, the system enters an incompressible Mott regime and the compressibility goes to zero. At the same time, the entropy develops sharp cusps as a function of density. The Mott gap can be obtained from studying the compressibility as a function of temperature.

In the incompressible Mott regime, the strong coupling expansions can be recast as a high temperature expansion for a spin model with SU(4) symmetry. 
The spin models at $\rho=1$ and $\rho=3$ belong to the fundamental representation of SU(4) symmetry with four states per site. The model at $\rho=2$ has six states per site and can be mapped to one with SO(6) symmetry \cite{zhang-mao}. The thermodynamic properties such as entropy and specific heat can be arranged in terms of two dimensionless parameters: $\beta t^2/U$ and $t/U$. At large $U/t$ the system turns into a Heisenberg model with $J$ set by $t^2/U$ whereas additional $t/U$ dependence reflects the presence of various multi-spin and multi-site interactions \cite{macdonald, delannoy, mila}. Antiferromagnetic ordering and correlations are captured by the $\beta t^2/U$ dependence whereas the $t/U$ terms can be used to study the breakdown of strong coupling expansions. 

The manner in which incompressibility sets in with increase in $\beta$ at various densities, and the shapes of chemical potential, inverse compressibility and entropy as a function of particle density carry potential lessons for magic angle graphene systems. While some features resemble those observed experimentally \cite{expts-a,expts-b,expts-c,expts-s1,expts-s2}, there are important differences, especially with respect to persistence of band features with temperature, which we will discuss in this paper.

\begin{figure}[htb!]
\includegraphics[width=\columnwidth]{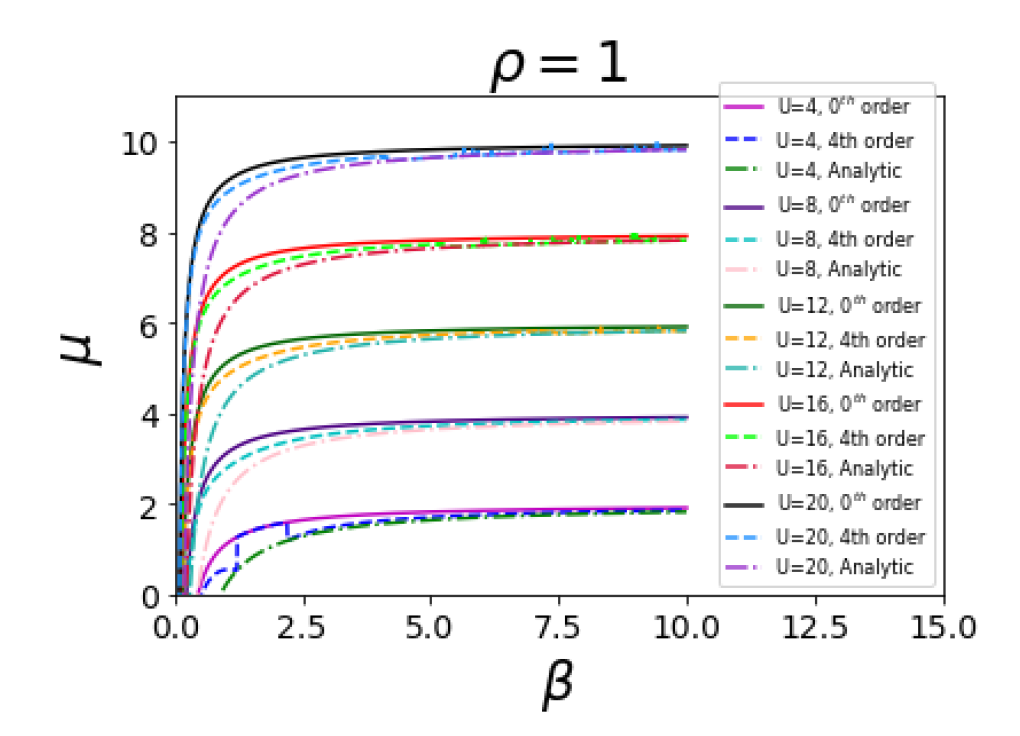}
\caption{Chemical potential $\mu$ versus inverse temperature $\beta$ for $\rho=1$ and several values of $U$. Zeroth order and fourth order calculations are shown along with an analytical asymptotic low temperature formula.}
\label{mu} 
\end{figure}

{\bf Model and Methods:} 
The SU(4) Hubbard model is defined by a Hamiltonian $H=H_0+V$, where the unperturbed part $H_0$ is the on-site term:
\begin{equation}
   H_0= U \sum_i {n_i (n_i-1)\over 2} -\mu \sum_i n_i,
\end{equation}
with $\mu$ the chemical potential and $n_i$ the total number operator on site $i$. 
The perturbation $V$ is the hopping term:
\begin{equation}
    V=-\sum_{i,j} t_{ij}   \sum_{\alpha=1}^4 (C_{i,\alpha}^\dagger C_{j,\alpha} + h.c.),
\end{equation}
where the sum $i,j$ runs over pairs of sites of a lattice and the sum over $\alpha$ runs over the $4$ species of Fermions. In this work, we will consider the nearest-neighbor hopping model on honeycomb lattice.

The unperturbed Hamiltonian has a particle-hole symmetry 
$n_i\to 4-n_i$, $\mu\to 3U-\mu$. The hopping Hamiltonian changes sign under particle-hole symmetry. Thus on a bipartite lattice where thermodynamic properties depend only on the absolute value of the hopping parameter $t$, the system has particle hole symmetry. 

Using the formalism of thermodynamic perturbation theory \cite{oitmaa-book, oitmaa2, ten-haaf, rrps1,rrps2},the logarithm of the grand partition function, per site, can be expanded as
\begin{equation}
    \frac{1}{N_s}\ln{Z}=\ln{z_0}+\sum_{G}L_G z_0^{-s} (\beta t)^r X_G(\zeta,\beta U),
    \label{graphs}
\end{equation}
where $N_s$ is number of sites in a large lattice, $z_0$ is the single-site partition function 
\begin{equation}
    z_0=1+4\zeta + 6\zeta^2 w 
    + 4\zeta^3 w^3 + \zeta^4w^6,
\end{equation}
with $\zeta=e^{\beta \mu}$, and the sum in Eq.~\ref{graphs} is over graphs denoted $G$. The graph $G$ has $s$ sites and $r$ bonds, and has Lattice Constant $L_G$. 
The weight-factor $X_G(\zeta,\beta U)$ is the reduced weight of the graph $G$ for $\ln{Z}$ \cite{oitmaa-book}.

\begin{figure}[htb!]
\includegraphics[width=\columnwidth]{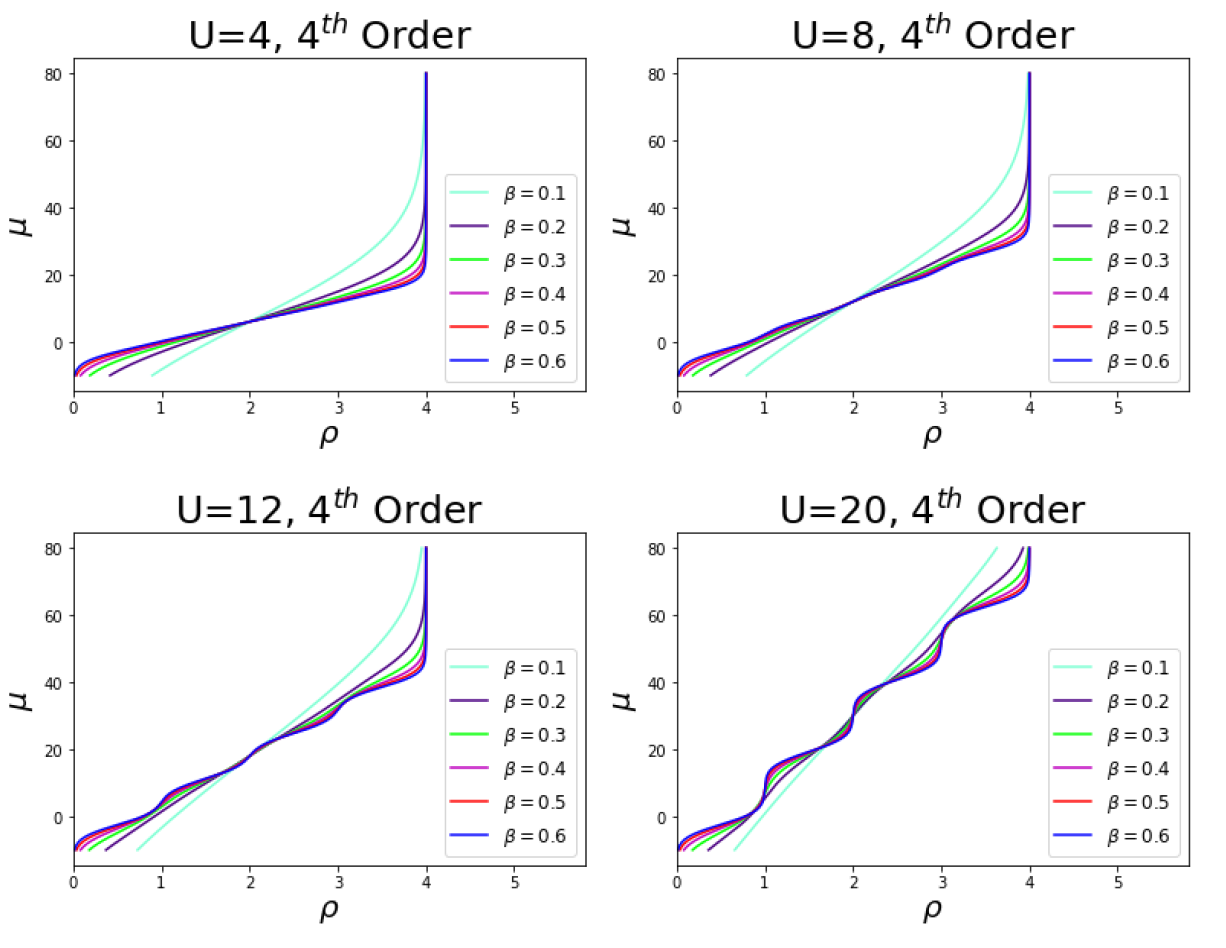}
\caption{Chemical potential $\mu$ versus density $\rho$ for several $U$ and $\beta$ values. As the temperature is lowered, the development of Mott behavior is indicated by rapid changes in chemical potential at integer densities. }
\label{muvsrho} 
\end{figure}

 The particle density (per site) can be obtained via the relation
\begin{equation}
    \rho=\frac{\zeta}{N_s}{\partial \over \partial \zeta} \ln{Z}.
\end{equation}
This relation can be inverted to obtain $\zeta$ or $\mu$ as a function of $\rho$ and $\beta$, which then allows one to obtain various properties at fixed particle density.

In Fig.~\ref{mu} we show the chemical potential $\mu$ as a function of temperature at particle density $\rho=1$ for different values of $U$. The solid lines are the calculation in the atomic limit (zeroth order), the dashed line in the fourth order, and the dashed-dotted lines is the asymptotic low temperature formula $\zeta^2=\frac{1}{6w}$, discussed more in a later section. 
We can see that the fourth order calculation lies between the zeroth order and the asymptotic expression. At low temperatures it becomes difficult to precisely locate the chemical potential numerically because the Mott plateaus have almost chemical potential independent density. 
Fortunately, the analytical expression can be used there.

The compressibility $K$, entropy $S$ and specific heat $C$ are obtained from the relations
\begin{equation}
    K= \big( \frac{\partial \rho}{\partial \mu} \big)_\beta,
\end{equation}
\begin{equation}
    S=-\beta({\partial \over \partial \beta}\ln{Z})_\zeta -\rho\ln{\zeta} + \ln{Z},
    \label{S}
\end{equation}
and,
\begin{equation}
 C=T\big (\frac{\partial S}{\partial T} \big )_\rho \ .
\end{equation}




We have carried out the perturbation theory to eighth order.
Up to fourth order, we evaluate the full traces and our perturbation theory is complete. For much of the temperature range studied, fourth order perturbation suffices and properties can be calculated accurately for arbitrary densities. Starting with sixth order the number of trace terms becomes too large to evaluate fully. In sixth order, we restrict trace calculations to at most 3 particles per site and in eighth order we restrict to at most two-particles per site. This restricted calculation is sufficient at lower temperatures ($w\to 0$ limit), where we particularly need higher orders, as discussed later. 


\begin{figure}[htb!]
\includegraphics[width=\columnwidth]{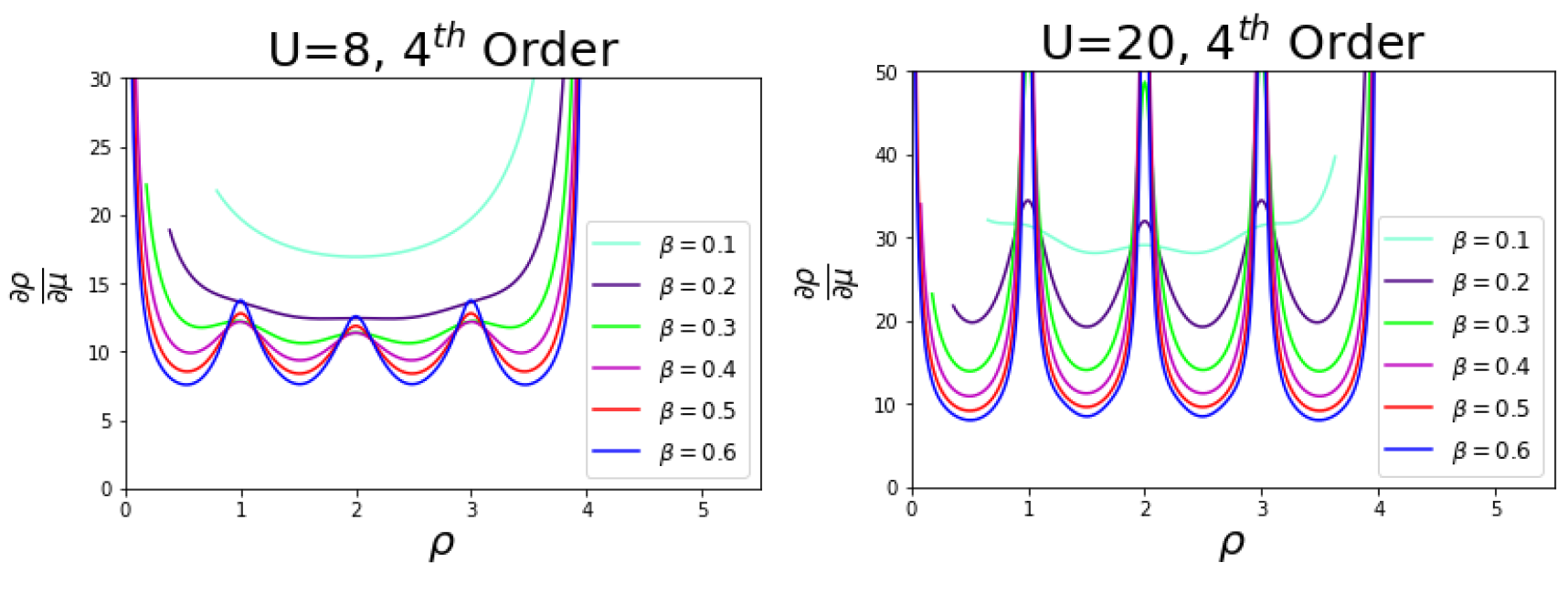}
\caption{Inverse Compressibility versus $\rho$ for $U/t=$ $8$ and $20$ and several $\beta$ values. Mott behavior is characterized by a sharp increase in inverse compressibility.}
\label{Kinvvsrho} 
\end{figure}

{\bf Chemical potential, Compressibility and Entropy:}
Numerical results for chemical potential $\mu$ as a function of particle density $\rho$ for various temperatures and U values are shown in Fig.~\ref{muvsrho}. We find that the results at all density are very well converged down to a temperature of approximately $T/t=1.5$. Below that temperature, the convergence away from integer densities starts to break down. 
Hence, the plots are shown up to $\beta=0.6$. At and near integer densities, the convergence of the expansion is set by $t^2/U$ and hence they remain convergent down to lower temperatures. The striking feature of the plot is the onset of Mott behavior around $T=U/2$ characterized by rapid rise in the chemical potential at integer densities. The larger $U$ system is deeper into the Mott phase and hence the changes in chemical potential are much sharper. Mott behavior can be seen even more clearly in Fig.~\ref{Kinvvsrho}, where we show the inverse compressibility vs density. The Mott phase is characterized by its incompressibility and hence the inverse compressibility shoots up and shows sharp spikes. Note that the spikes are symmetric around the peak. 

\begin{figure}[htb!]
\includegraphics[width=\columnwidth]{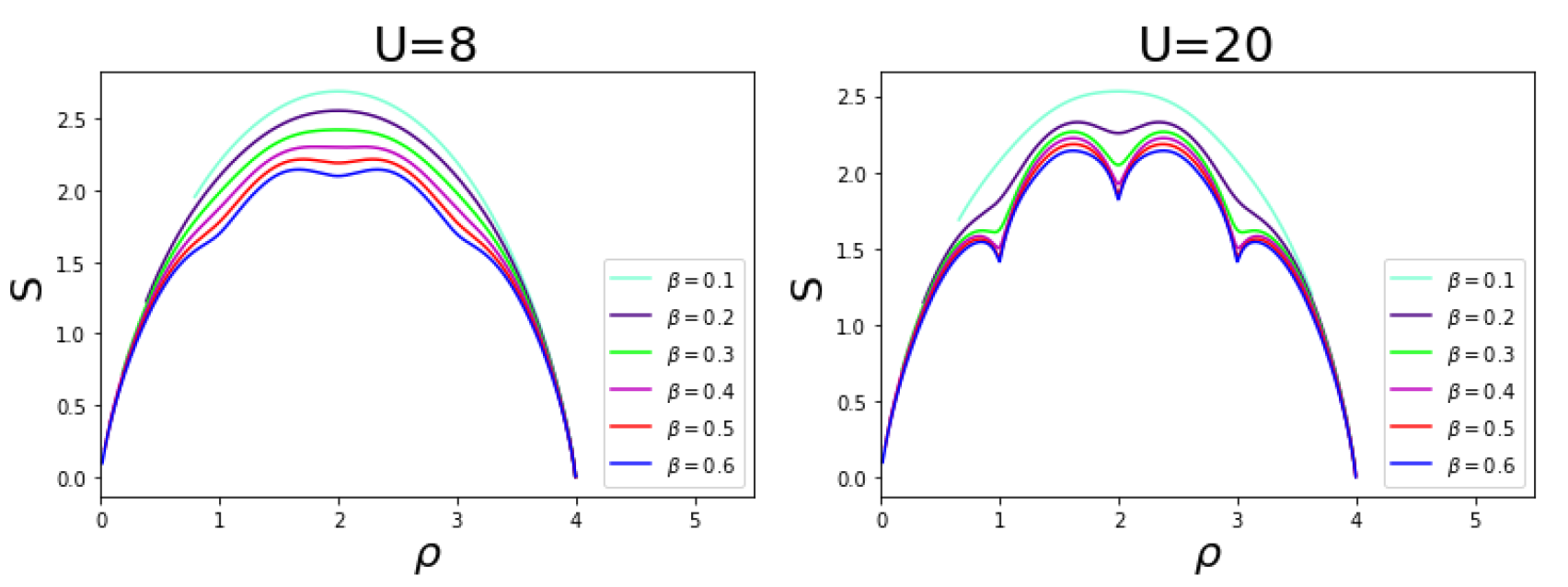}
\caption{Entropy as a function of density for several values of $\beta$ and $U/t=$ $8$ and $20$. As the system enters the strongly
correlated regime sharp cusps develop at integer densities.}
\label{Svsrho} 
\end{figure}

In Fig.~\ref{Svsrho}, we show the entropy as a function of density. The onset of Mott behavior is characterized
by the development of a sharp cusp in the entropy with a minima at integer densities. This is because at small deviation from integer
densities and at temperatures much larger than the exchange constant, the system maps on to the high temperature
limit of the $t-J$ model \cite{rrps1,glenister,rigol,putikka,pryadko}. At these temperatures, the motion of charge degrees of freedom is incoherent and corresponds to a dilute gas with an entropy that varies as $- \delta\ln{\delta}$, where $\delta$ is the deviation in density from integer filling.

%

{\bf The $w\to 0$ limit:}
At low temperatures, $w=\exp{-\beta U}$ becomes exponentially small and some terms dominate the expansion. At integer densities, the system is dominated by a single occupancy value at each site. Away from integer densities, the system is dominated by only two occupancy values. The remaining terms become exponentially small. To see this, and the range of $\zeta$ or chemical potential $\mu$ in each case, we consider the unperturbed atomic limit.

Particle density in the atomic limit is:
\begin{equation}
    \rho_0 = \frac{1}{z_0} (4\zeta +12 \zeta^2 w + 12 \zeta^3 w^3 + 4 \zeta^4 w^6)
    \label{density0}
\end{equation}
The cases $\rho=0$ and $\rho=4$ are trivial. We focus on remaining integer densities. Setting $\rho=1$ and dropping exponentially small terms
one obtains:
\begin{equation}
    \zeta^2 =\frac{1}{6w} + \mathcal{O}(w).
    \label{rho1}
\end{equation}
This relation is exact at low temperatures for $\rho=1$ provided the system remains in the Mott phase, that is, U is not so small that there is a transition away from the insulating phase.
The single site partition function is also dominated by a single term
\begin{equation}
    z_0=4\zeta +\mathcal{O}(\sqrt{w}).
\end{equation}
Or,
\begin{equation}
    \frac{\zeta}{z_0}=\frac{1}{4} + \mathcal{O}(\sqrt{w})
\end{equation}
For each graph, in the $w\to 0$ limit, $X_G$ is also dominated by only select terms. For a graph with $s$ sites, there is a $z_0^s$ factor in the denominator and the numerator has a maximum power of $\zeta^s$ without any double occupancy. Thus, only the $\zeta^s$ terms survive in this limit. 
In other words, graph by graph, only those terms survive which have exactly one particle at every site in the unperturbed limit.

Setting $\rho=2$ in Eq.~\ref{density0} we get
\begin{equation}
    \zeta^2 =\frac{1}{w^3}.
    \label{rho2}
\end{equation}
This relation, which implies $\mu=3U/2$, is exact by particle hole symmetry.
The partition function, up to terms which are relatively exponentially small, becomes:
\begin{equation}
    z_0=6\zeta^2 w.
\end{equation}
Or,
\begin{equation}
    \frac{\zeta^2}{z_0}=\frac{1}{6w}.
\end{equation}
Once again, contribution from each graph is dominated by terms that have exactly two particles on every site. Other terms are relatively exponentially small.

Setting $\rho=3$ in Eq.~\ref{density0} and keeping only the exponentially largest terms we get $\zeta^2 =\frac{1}{6w^5}$.
The partition function, up to terms which are relatively exponentially small, becomes $ z_0=4\zeta^3 w^3$.
Or,
\begin{equation}
    \frac{\zeta^3}{z_0}=\frac{1}{4w^3}.
\end{equation}
Once again, contribution from each graph is dominated by terms that have exactly three particles on every site. Other terms are relatively exponentially small.

At densities between two integer densities, the fugacity takes values between two commensurate ones and each site can have only one of two occupations. Thus, restricting trace calculations to terms with up to $2$ particles per site suffices to get the exponentially largest terms as long as particle density is less than or equal to two. Furthermore, using particle-hole symmetry one can also obtain properties at densities larger than two, so that all densities can still be accurately obtained.

As $w\to 0$, the expansions at integer densities turn into high temperature expansions for a generalized spin model. We can rearrange this expansion in powers of $x=\beta t^2/U$ and $y=(t/U)^2$.
At $\rho=1$, $\ln{(Z/4)}$ has expansion
\begin{equation}
\begin{split}
    & 2.25x (1 -\frac{4}{3}y + 15.81887 y^2
    -429.6101 y^3 + \ldots) \\
    &+ 2.8125 x^2 (1 -\frac{16}{3} y + 66.49491 y^2 + \ldots)\\
    &+ 0.9375 x^3 (1 - 0.8 y + \ldots ) - 0.00234375 x^4 (1 + \ldots)
    \end{split}
    \label{rho1-h}
\end{equation}
While at $\rho=2$, $\ln{(Z/6)}$ has expansion
\begin{equation}
\begin{split}
    & 3x (1 -1.833333y + 34.18148 y^2
    - 1227.889 y^3 + \ldots) \\
    &+ 5 x^2 (1 -9.516667 y + 182.3093 y^2 + \ldots)\\
    &+ \frac{20}{3} x^3 (1 - 9.866667 y + \ldots ) +\frac{26}{9} x^4 (1 + \ldots)
    \end{split}
    \label{rho2-h}
\end{equation}
The $y\to 0$ limit corresponds to the Heisenberg model and the properties only depend on $\beta t^2/U$. As long as the expansions in $t/U$ converge, the system remains an incompressible Mott Insulator. 

Series in $y$ are too short to determine the location of metal-insulator transition. But, taking the nth root of the absolute value of the coefficients of various $y^n$ terms suggest a convergence radius in $(t/U)^2$ of approximately  $0.1$ or $U/t\approx 3-4$.  Note that the critical $U/t$ need not be the same at different densities. Also, because the series in $y$ are alternating, we cannot rule out a much smaller critical $U/t$ on the real axis. These results are consistent with a previous study \cite{fu3}, which reported a critical $U/t$ in the range of $2.5$ to $3$.

\begin{figure}[htb!]
\includegraphics[width=\columnwidth]{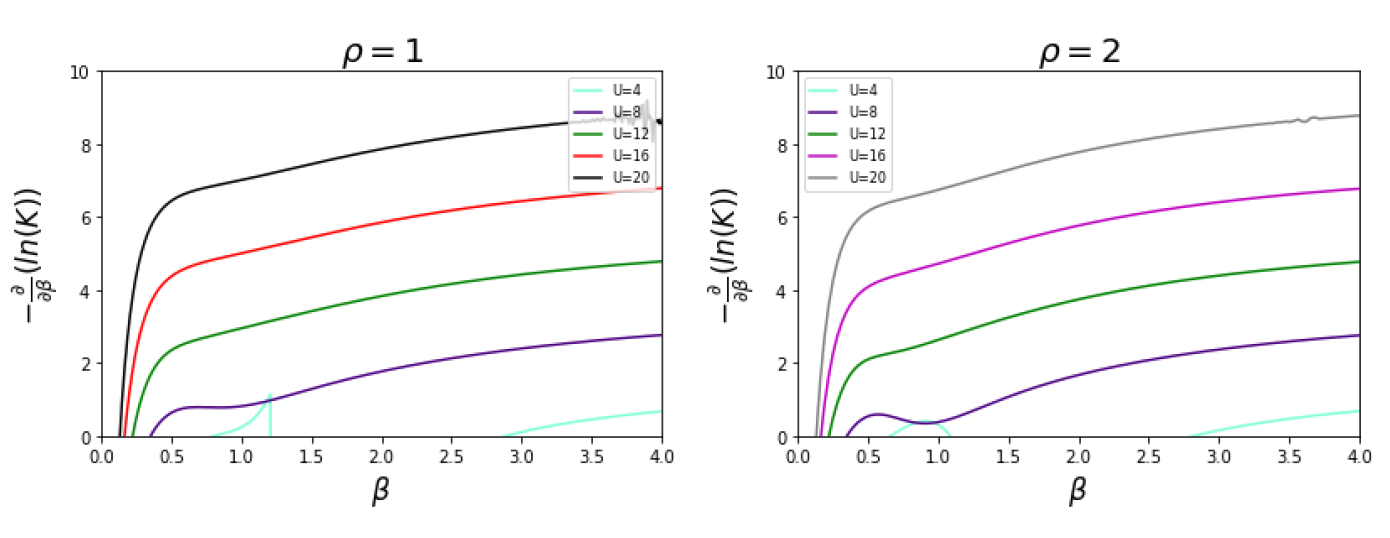}
\caption{Effective Mott Gap defined as $-\frac{d \ln{K}}{d\beta}$, versus temperature for $\rho=1$ and $\rho=2$ for several values of $U$.}
\label{Kvsbeta} 
\end{figure}

{\bf Determining the Mott gap:}
We define the effective Mott Gap from the behavior of compressibility as $K\sim \exp{-\Delta/T}$. The effective Mott gap at inverse temperature $\beta$ is defined as:
\begin{equation}
    \Delta = - \frac{d \ln{K}}{d\beta}
\end{equation}
Fig.~\ref{Kvsbeta} shows the effective Mott gap as a function of $\beta$ for several $U$ values. In finite order of perturbation theory, at asymptotically low temperatures the Mott gap thus defined can be shown to go back to $U/2$, the unperturbed value. The reason is simply that $\exp{(-\beta U/2)} (\beta t)^n$ goes to zero as $n \to \infty$. Hence, keeping $n$ finite, as $\beta\to\infty$ all perturbative terms vanish. In nth order, convergence should extend up to a $\beta$ value that increases as $n$. For our calculation, we thus need to stay at a low but finite temperature. We expect the Mott gap to plateau close to the true answer before drifting back to the unperturbed value as $T\to 0$. It is, however, difficult to precisely pin down the gap from the short series. It seems apparent that the gap becomes small by $U/t=4$.
At smaller $U$, $U$-dependence of the gap may take a concave shape so that a small gap may persist to much smaller $U$ values.


\begin{figure}[htb!]
\includegraphics[width=\columnwidth]{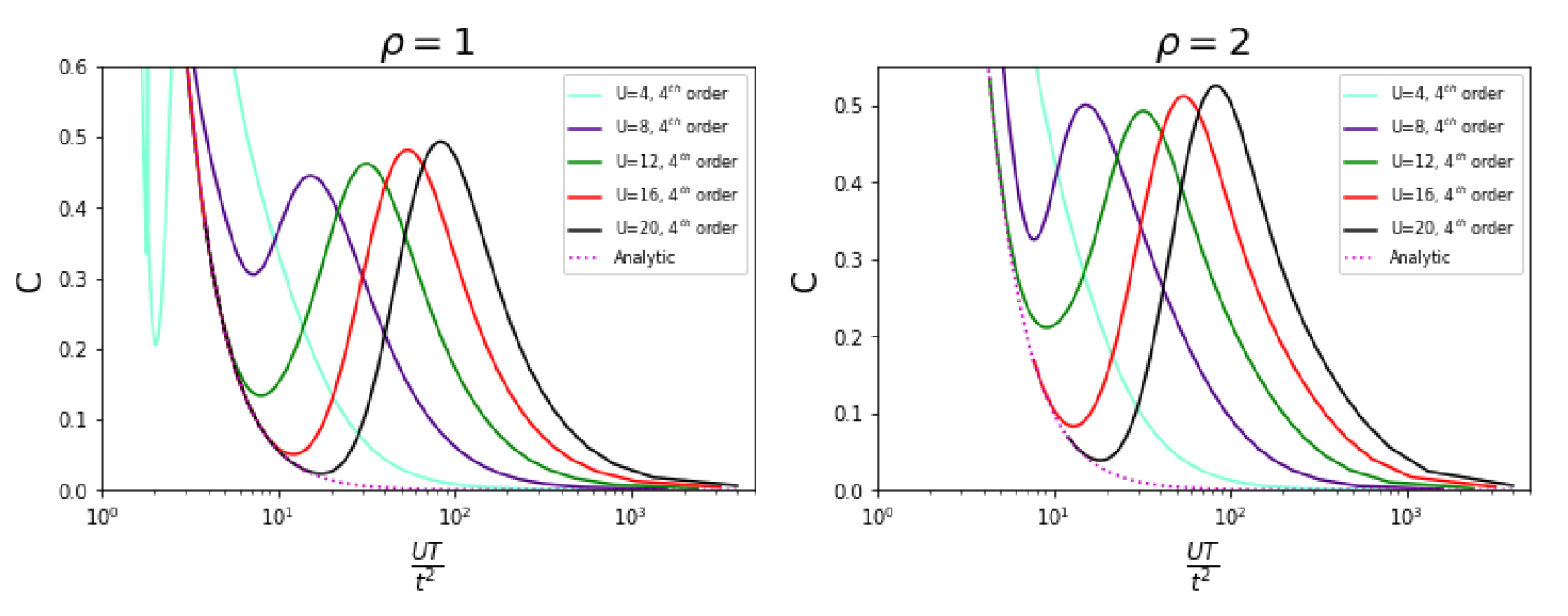}
\caption{Specific Heat (C) as a function of temperature for $\rho=1$ and $\rho=2$ for several values of $U$. The leading order Heisenberg model results for specific heat are shown as dotted lines.}
\label{SC} 
\end{figure}


{\bf Specific heat in the insulating phase:}
In Fig.~\ref{SC} specific heat plots for $\rho=1$ and $\rho=2$ at various $U$ values as a function of temperature are shown. The temperature axis has been scaled by $t^2/U$ to match with the large-U exchange constant. The high temperature peak in specific heat corresponds to the transition from temperatures of order $U$ to a strongly correlated regime at $T<<U$. For $U\ge 8$, the entropy first plateaus at $ln{4}$ and $\ln{6}$ for $\rho=1$ and $\rho=2$ respectively, corresponding to the high temperature limit of the spin model \cite{rrps1}. Subsequently, the entropy decreases with lowering of temperature showing the development of spin correlations. The lower temperature rise of the specific heat is related to the development of spin correlations in the system.

We have also shown in Fig.~\ref{SC}, the high temperature large-U limit of the spin models for the specific heat. It is clear that the spin model captures the numerical variations in the quantities for $U\ge 8$ quite well. But, there are significant corrections for $U/t=4$.

{\bf Magic angle systems:} 
In the Twisted Bilayer Graphene and similar systems near magic angles the bandwidth and effective $U$ have both been estimated to be of order $10$ meV \cite{expts-a}. That means the hopping parameter $t$ is of order $10$K and $U$ is of order $100$K. Thus, our calculations could be relevant above a temperature of a few K at integer densities and above about $10$K at all densities.

There have been several measurements of chemical potential, compressibility and entropy as a function of density in these systems some going to temperatures up to $50$K or higher \cite{expts-a, expts-b,expts-c, expts-s1, expts-s2}. Some features of the experiments are clearly captured by the simple Hubbard model. As temperature is lowered, there occurs a rapid rise in  chemical potential at characteristic densities and correspondingly the inverse compressibility shows sharp spikes. However, there are key differences:

1. Mott behavior at $T>1.5t$ is only seen in our study at integer densities per site not at integer densities per translational unit cell as seen in experiments. The latter would include half integer densities per site, implying twice as many Mott plateaus. 

2. The entropy is a minima at integer densities in our studies with sharp cusps, whereas it is a maxima in experiments.

3. No saw-tooth like asymmetry is seen in the inverse compressibility spikes \cite{expts-b}, though this maybe related to the particle hole symmetry in the  nearest-neighbor hopping model.

Although our model has only nearest-neighbor hopping, results 1 and 2 above should be valid even with more complex hoppings. At temperatures above $t$ and $U$ not too small, incompressibility in the Hubbard model only arises when it can be traced back to the atomic limit, i.e., at integer density per site. Adding further neighbor Coulomb repulsion can cause additional Mott plateaus at half-integer densities, but that would be related to charge density order for which there is no experimental evidence. In fact, there is a band-based argument for insulators at half integer densities.
The band structure of the honeycomb system can be regarded as two bands, one below the other in energy, joined together at the Dirac points \cite{expts-b}. Thus, additional incompressibility at half integer density per site is related to integer filling of one of the two bands.

Similarly, hopping of carriers will normally be incoherent at these high temperatures in the Hubbard model and the mobile particle entropy will be that of an ideal gas, regardless of hopping details. This is what causes the entropy function to be a minima at integer densities and have sharp cusps. On the other hand, the kinetic entropy is presumably already quenched even at these temperatures in experiments and the system has turned into a fermi liquid.

We believe these differences point to an important property of magic angle flat-band systems, namely that these flat bands are derived from a much wider band and hence have a much larger energy scale behind them and thus can persist over a larger temperature scale. Persistence of low energy Dirac features to high temperatures has also been emphasized by Zondiner et al \cite{expts-b}.


{\bf Discussions and Conclusions:}
We have studied finite temperature properties of the SU(4) Hubbard model on the honeycomb lattice using strong coupling expansions. At integer densities, when U is not too small, the system becomes an incompressible Mott Insulator. This can be seen by examining the density versus chemical potential, which shows the development of sharp plateaus. When this happens, the compressibility becomes exponentially small and the entropy develops cusps at integer densities. The system can be mapped to a spin model with only virtual charge fluctuations. These expansions converge extremely well for $U/t>8$ and possibly down to $U/t\approx 3-4$. 

Strong coupling expansion is particularly useful in studying the temperature dependence of properties like compressibility and entropy. 
For a more quantitative comparison with experiments on magic angle flat-band materials, it may be useful to extend these studies to include more realistic band structures with many hopping parameters. 
Furthermore, including smaller terms that break SU(N) symmetry would allow one to study various symmetry breaking transitions in these systems. However, our study suggests that some aspects of the physics of magic angle systems may not be captured by considering a lattice Hubbard model of just the flat bands.

{Acknowledgement:} This work is supported in part by the US National Science Foundation grant DMR-1855111. One of the authors (JO) acknowledges computing support provided by the Australian National Computation Infrastructure (NCI) Program.

\end{document}